# Quantum materials QED with van der Waals crystals

**Authors:** Suheng Xu[1*], Brian Vermilyea[2], Ran Jing[3,4], Lukas Wehmeier[5], Jihoon Park[3], Rafael A. Mayer[3], Xinzhong Chen[1,3], Samuel L. Moore[1], Fuyang Tay[1,6], Boyi Zhou[1,3], Wenjun Zheng[3], Dihao Sun[1], Birui Yang[1], Cory R. Dean[1], Daniel Cazorla[3], Xu Du[3], Milan Delor[6], Andrew J. Millis[1], Angel Rubio[7,8,9], Michael M. Fogler[2], Mengkun Liu[3,5]*, D.N. Basov[1]*

**Affiliations:**

[1]Department of Physics, Columbia University; New York, New York 10027, USA.

[2]Department of Physics, University of California San Diego; La Jolla, California 92093-0319, USA.

[3]Department of Physics and Astronomy, Stony Brook University; Stony Brook, New York 11794, USA.

[4]Condensed Matter Physics and Materials Science Division, Brookhaven National Laboratory; Upton, New York 11973, USA.

[5]National Synchrotron Light Source II, Brookhaven National Laboratory; Upton, New York 11973, USA.

[6]Department of Chemistry, Columbia University; New York, New York 10027, USA.

[7]Theory Department, Max Planck Institute for Structure and Dynamics of Matter and Center of Free-Electron Laser Science, 22761 Hamburg, Germany

[8]Center for Computational Quantum Physics, The Flatiron Institute, 162 5th Avenue, New York, New York 10010, USA

[9]Nano-Bio Spectroscopy Group, Universidad del País Vasco UPV/EHU, San Sebastián 20018, Spain

*Corresponding authors. S. Xu: sx2277@columbia.edu, M.K. Liu: mengkun.liu@stonybrook.edu, D.N. Basov: db3056@columbia.edu

# Abstract

**A central goal of the emerging field of materials QED is to harness subwavelength electromagnetic confinement in engineered cavities to tailor light–matter interactions. Here, we demonstrate that van der Waals multilayer cavities composed of stacked graphene and hexagonal boron nitride (hBN) provide unprecedented control over hybridization between their electromagnetic resonances. By magnetically tuning graphene inter-Landau-level transitions (ILTs) into resonance with hBN phonons, we witness broadband, mode-selective strong coupling in the mid-infrared. Our quantum-material QED framework shows that cavity geometry and layer placement jointly control both coupling strength and modal selectivity. For experimentally realistic cavity parameters, the predicted mode splitting exceeds the width of the hBN Reststrahlen band, providing a direct route to the ultrastrong coupling regime. Together, these results establish van der Waals multilayer cavities as a scalable platform for tunable multimode QED and broaden the design space for reconfigurable quantum materials.**

# Introduction

Cavity quantum electrodynamics (c-QED) describes the interaction between matter and quantized electromagnetic fields [1–3]. Conventional realizations of c-QED rely on engineered cavities, such as mirror pairs or Fabry–Pérot resonators, coupled to atoms, molecules or quantum-dot emitters. In quantum materials, strong hybridization between photonic modes and intrinsic matter resonances has been theoretically predicted to modify phase transitions and ground-state properties of materials embedded in cavities [1–7]. Recent experimental observations of cavity-modified material properties [7–16] have established QED engineering as a rapidly developing frontier at the interface of quantum optics and condensed matter physics. A central challenge is therefore to identify cavity architectures in which material excitations themselves actively shape the electromagnetic environment.

Two-dimensional (2D) quantum materials and van der Waals (vdW) heterostructures provide a powerful platform for realizing QED phenomena. These materials host prominent dipole-active collective modes arising from electronic, plasmonic, excitonic, and phononic resonances [17–27]. When vdW crystals are vertically stacked, collective modes in adjacent layers hybridize, giving rise to emergent optical responses that cannot be understood as a simple superposition of their constituents [23,28–32]. A representative example is plasmon-phonon coupling at graphene/hBN interfaces, extensively characterized by nano-optical studies [30,31,33–35]. Extending beyond single interfaces, multilayer graphene-dielectric superlattices have been proposed as hyperbolic plasmonic media supporting collective electromagnetic effects [36–39]. Automated van der Waals assembly has produced superlattices comprising 29 alternating graphene and hBN layers, demonstrating the feasibility of highly repeated stacks [40]. Building on this foundation, we show that collective hybridization across layers enables layer-resolved control of optical field profiles and, crucially, opens pathways to the ultrastrong coupling regime ($\frac{\Delta}{2} \gtrsim 0.1\omega_0$), where mode splitting (Δ) exceeds 20% of the bare resonance frequency $\omega_0$ [3]. Multilayer vdW cavities therefore offer functionalities fundamentally inaccessible in single-interface systems.

Realizing controllable multilayer hybridization experimentally, however, requires a tunable mechanism to activate electromagnetic response in charge-neutral graphene, which does not itself support strong plasmonic resonances. Common approaches to initiating plasmons such as electric gating or chemical doping are prohibitively difficult to implement in multilayer cavities. Magnetic fields provide a powerful alternative: Landau-level (LL) transitions in charge-neutral graphene generate strong mid-infrared resonances without external doping. Recent nano-magneto-optical experiments [41–43] have demonstrated that these magnetically induced resonances enable robust plasmon-phonon hybridization in a multilayer structure. Magnetic tuning thus converts the vdW multilayers into reconfigurable, deeply subwavelength cavities, in which Landau-level transitions and phonon resonances can be brought into and out of resonance through applied field control or layer positioning.

Fig. 1(a) illustrates a representative graphene/hBN multilayer cavity (MC) composed of four unit cells with a total thickness of 70 nm, denoted ($d = 70nm, N = 4$). Each unit cell comprises a graphene sheet and an hBN spacer, with additional hBN cladding at the outer surface. The complete stacking sequence is hBN/graphene/hBN/…/graphene/hBN. Owing to its strong optical anisotropy, hBN supports deeply subwavelength hyperbolic phonon resonances [31,44–48]. These hyperbolic modes resemble Fabry–Pérot resonances but operate far below the diffraction limit [12]. Previous studies have shown that such hBN-based cavities enable extreme field confinement and enhanced vacuum fluctuations, supporting strong light-matter coupling [28,45] and even modifying the ground state of a superconductor embedded in a cavity [28,45]. Here, we leverage these cavity properties to construct a multilayer quantum-material platform in which hyperbolic phonons in hBN strongly hybridize with electronic Landau-level resonances in graphene. When graphene layers are embedded within hBN slabs, inter-LL transitions (ILTs) couple to hBN phonon modes [42,49]. As shown in Fig. 1b, at zero detuning between the ILTs and the hyperbolic phonon resonance, the system exhibits a characteristic avoided crossing, forming upper and lower hybrid branches with comparable electronic ILT and phonon character.

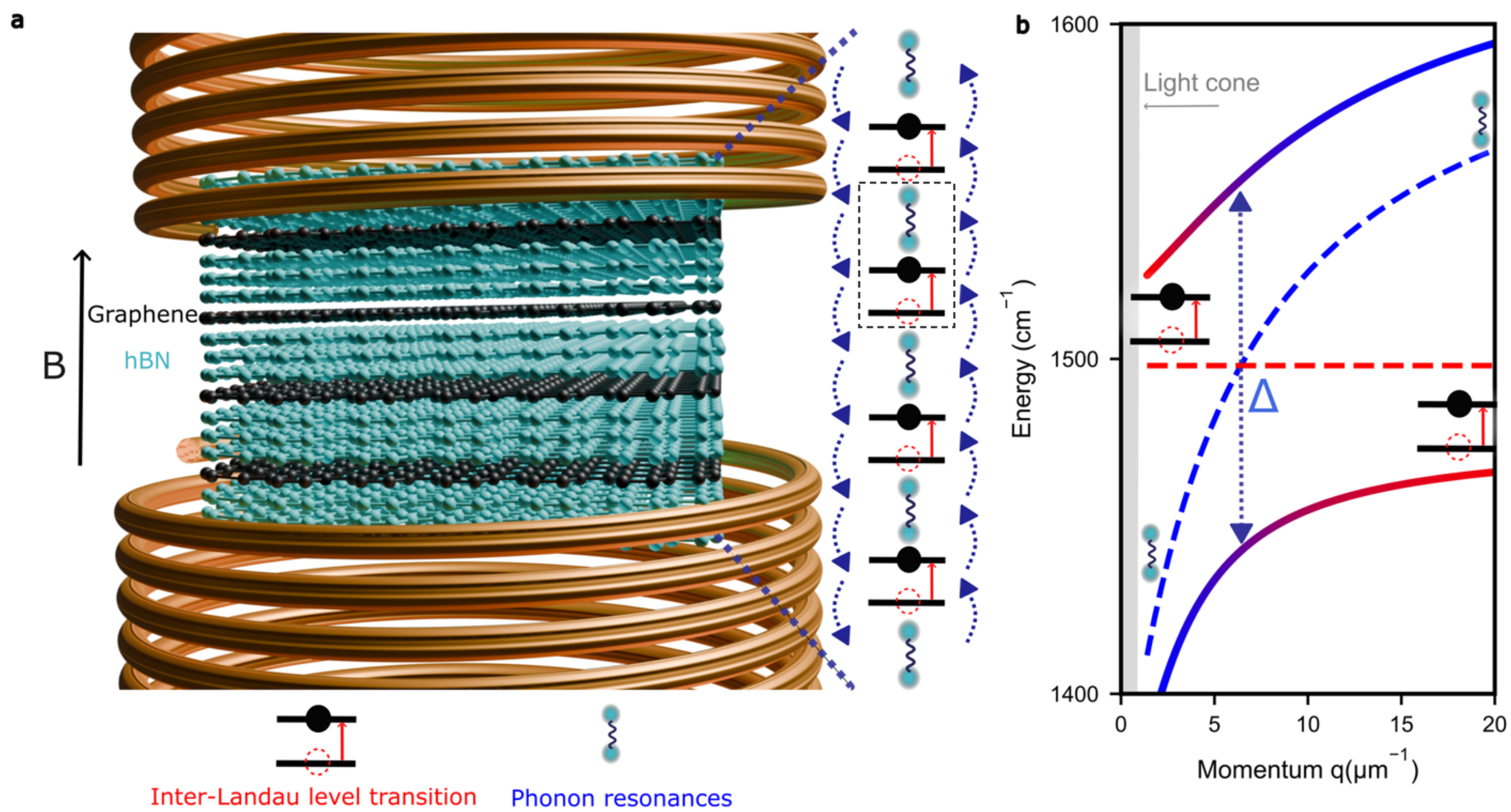

Figure 1. Quantum-material QED in a graphene/hBN multilayer cavity. (a) Schematics of a graphene/hBN multilayer cavity composed of four unit cells under a perpendicular magnetic field. The black and blue atomic layers represent graphene and hBN, supporting inter-Landau-level transitions (ILTs) and hyperbolic phonon resonances, respectively. The hyperbolic phonon resonances in hBN couple strongly to graphene ILTs through long-range Coulomb interactions across adjacent layers. The dashed rectangle in the right inset marks one graphene/hBN repeat. (b) Mode hybridization and avoided crossing in a 70-nm-thick MC with four unit cells. The red and blue dashed curves show the dispersions of the uncoupled graphene ILT and the hBN hyperbolic phonon cavity mode, respectively. The solid curves represent the hybridized modes, with color indicating the relative contributions from the ILT and phonon components. The gray shaded area denotes the light cone, highlighting the subdiffractional nature of mode coupling in the heterostructure cavity.

The avoided crossing and mode hybridization in graphene/hBN MCs result from electromagnetic coupling between internal degrees of freedom in layered van der Waals heterostructures. Notably, these effects do not require external optical illumination; instead, they originate from strongly confined near-field electromagnetic modes intrinsic to the materials themselves. We refer to this regime as quantum materials QED, emphasizing its origin in the material's internal electromagnetic properties. In the following sections, we analyze quantum materials QED in graphene/hBN multilayer cavities using a fully quantized framework. We compute the spatial profile of hyperbolic cavity modes, enabling direct access to mode-resolved and depth-dependent mode hybridization. We demonstrate that the vertical placement of graphene layers within the hBN stack controls both the magnitude and character of mode coupling, producing tunable splitting and nontrivial mode structures. At large in-plane momenta, the MC can enter the ultrastrong coupling regime, opening opportunities to explore ground-state modifications and potential superradiant phase transitions [50]. These findings underscore the importance of spatially resolved quantum treatments in vdW polaritonics and establish MCs as a promising platform for tunable, multimode QED in the mid-infrared regime.

## Quantum electrodynamics of graphene/hBN multilayer cavities

To elucidate the cavity-induced effect in the MC in Fig. 1a, we develop a quantum material QED framework that explicitly incorporates discrete layer geometry, magnetically tunable ILTs, and full electromagnetic boundary conditions. The confined electromagnetic field is quantized using a scalar potential that satisfies exact boundary conditions at all interfaces (Supplemental Material). This framework resolves the z-dependent field distribution and electrostatic coupling between hyperbolic phonons and Landau-level excitations with layer-by-layer precision.

The QED Hamiltonian is derived in the near-field regime, where the frequency $\omega$ and in-plane momentum $q$ satisfy $q \gg \omega/c$. We adopt a spatially local continuum model for hBN, which accurately describes low-order phonon-polariton dispersion in slabs as thin as three atomic layers [44]. We also neglect the direct electrostatic coupling between different graphene layers. When an hBN phonon mode is tuned close to resonance with an ILT, we apply the rotating-wave approximation (RWA) and retain only the resonant interaction terms. Under these conditions, the Hamiltonian takes the form:

$$H = \sum_{\mathbf{q}} \Big[ \hbar\omega_{l,\mathbf{q}} a_{l,\mathbf{q}}^{\dagger} a_{l,\mathbf{q}} + \hbar\omega_{\mathrm{ILT}} \sum_{i=1}^{N} \sum_{j=1}^{2} b_{i,j,\mathbf{q}}^{\dagger} b_{i,j,\mathbf{q}} + \sum_{i=1}^{N} \sum_{j=1}^{2} e(E_{l,q}(z_i)\mu_{\mathrm{ILT}} a_{l,\mathbf{q}} b_{i,j,\mathbf{q}}^{\dagger} + \mathrm{H.c.}) \Big]. \quad (1)$$

Here $a_{l,\mathbf{q}}^{\dagger}$ denotes the creation operator for the l-th order hBN phonon cavity mode branch with frequency $\omega_{l,\mathbf{q}}$. The operators $b_{i,1,\mathbf{q}}^{\dagger}$ and $b_{i,2,\mathbf{q}}^{\dagger}$ are creation operators for two degenerate optically allowed ILTs (or magnetoexcitons) in graphene, each with transition frequency $\omega_{\mathrm{ILT}}$. These neutral particle–hole excitations couple to the longitudinal in-plane field of the hBN modes, connecting our framework to the established descriptions of magnetoplasmon–phonon coupling in two-dimensional electron gases [51,52]. The present model neglects many-body corrections to the ILT energies and therefore treats bare ILTs rather than Coulomb-dressed magnetoplasmon eigenmodes. A detailed derivation of Eq. (1) is provided in the Supplemental Material.

The last term in Eq. (1) describes a Fröhlich-type electron-phonon coupling. Here, $E_{l,q}(z)$ denotes the in-plane electric-field amplitude of the l-th hBN phonon cavity mode at in-plane wavevector q and vertical position $z$ in the MC, while $\mu_{\mathrm{ILT}}$ denotes the ILT dipole moment. The field amplitude, and hence the coupling strength, depends on q even when the ILT dipole is evaluated in the long-wavelength limit. The variable $z_i$ denotes the vertical position of the i-th graphene layer within the hBN slab. The explicit z-dependence enables geometry-selective control of the hybridization strength by positioning graphene layers within the cavity-mode field profile.

Figure 2(a,b) shows the calculated dispersion $\omega_{l,q}$ and corresponding normalized in-plane electric field profile $\mathcal{E}_{l,q}(z)$ of the l-th order hyperbolic phonon cavity mode in a 70-nm-thick hBN slab [31,44–46,53,54] (see Supplemental Material). The displayed profiles are evaluated at 1500 cm−1 using the corresponding wavevector for each branch. Higher-order modes occur at larger in-plane momenta and exhibit stronger field confinement, as evidenced by the rapid decay of the electric field outside the hBN slab (Fig. 2b). The l-th order mode has l nodes in its field profile within the hBN volume.

The Landau level (LL) energies in graphene are given by $E_n = \operatorname{sgn}(n)\sqrt{2|n||B|\hbar e v_f}$, where $n$ is the LL index, $B$ is the magnetic field, $\hbar$ is the reduced Planck constant, $e$ is the elementary charge, and $v_f$ is the Fermi velocity [41,55,56]. At $B$ =3.2 T, the $-2$→1 and $-1$→2 LL transitions produce an infrared-active resonance at $\omega_{LL}$= 1500 cm$^{-1}$, which lies within the bandwidth of the hyperbolic phonon cavity mode. The interaction between these ILTs and the hyperbolic phonon is described by the QED Hamiltonian introduced above.

We diagonalize the Hamiltonian in Eq. (1) to obtain the eigenenergies and eigenstates for a representative structure (d = 70 nm, N= 1), shown in Fig. 2(c-d). Assuming a momentum-independent ILT frequency and focusing on the fundamental (l = 0) phonon cavity mode, the coupling produces upper ($|+\rangle$) and lower ($|-\rangle$) hybrid branches separated by an anticrossing gap at the zero-detuning momentum $q_l$ defined by $\omega_{l,q_l} = \omega_{\mathrm{ILT}}$, as shown in Fig. 2c. Here, $|PP\rangle$ and $|LL\rangle$ denote the eigenstates of the hBN phonon mode and the graphene ILTs, respectively. The hybridized states $|\pm\rangle = \alpha|PP\rangle + \beta|LL\rangle$ exhibit momentum-dependent composition. At zero detuning ($q = 6.5\ \mu m^{-1}$), the mode contains equal phonon and ILT character. Such mode splitting has been experimentally observed using magneto-scanning near-field optical microscopy (m-SNOM) in the ($d = 71\ nm, N = 1$) MC [42,43], where strong hybridization produces a splitting of up to 50 cm$^{-1}$ within the hBN Reststrahlen band (1355 cm$^{-1}$ to 1605 cm$^{-1}$).

The zero-detuning mode splitting $\Delta_l$ between the ILTs and the l-th hBN phonon cavity mode in an MC with N>1 can be written in a form analogous to that in cavity QED:

$$\Delta_{\mathrm{l}} = \mu\left[\frac{\omega}{2\hbar d_{\mathrm{eff}}}\sum_{i=1}^{N}\mathcal{E}_{l,q_l}^2(z_i)\right]^{\frac{1}{2}}, \tag{2}$$

where $\omega$ is the common hBN-mode and ILT frequency at zero detuning, $\mu$ is the dipole moment per unit length for the ILTs, $d_{\mathrm{eff}} \propto d$ is the effective confinement length of the hBN phonon mode, and $\mathcal{E}_{l,q_l}(z)$ is the dimensionless in-plane field profile evaluated at the resonant wavevector, normalized such that $\frac{1}{d}\int_0^d \mathcal{E}_{l,q_l}{}^2(z)dz = 1$. All quantities are defined in the Supplemental Material. For coupling to the fundamental (l = 0) cavity modes, the mode splitting $\Delta_0$ depends weakly on graphene layer positioning because the electric field is nearly uniform throughout the hBN slab. In the large $N$ limit, we have $\sum_{i=1}^{N}\mathcal{E}_{l,q_l}{}^2(z_i) \approx N$, and the splitting reduces to

$$\Delta_0 = \mu\sqrt{\frac{\omega N}{2\hbar d_{\mathrm{eff}}}}. \tag{3}$$

This scaling resembles the light-matter coupling in cavity QED, where the effective cavity mode volume is replaced by the inverse graphene spacing $d_{eff}/N$. In this regime, the hBN phonon couples to the ILTs in all graphene layers, resulting in a collective enhancement of the coupling strength, analogous to recent work on exploiting multiple emitters or modes to boost the effective coupling strength [57–59]. The enhancement arises from the summed oscillator strengths of layer-resolved ILTs coupled to a common cavity mode, providing a scalable mechanism for strengthening light-matter interaction across the multilayer stack.

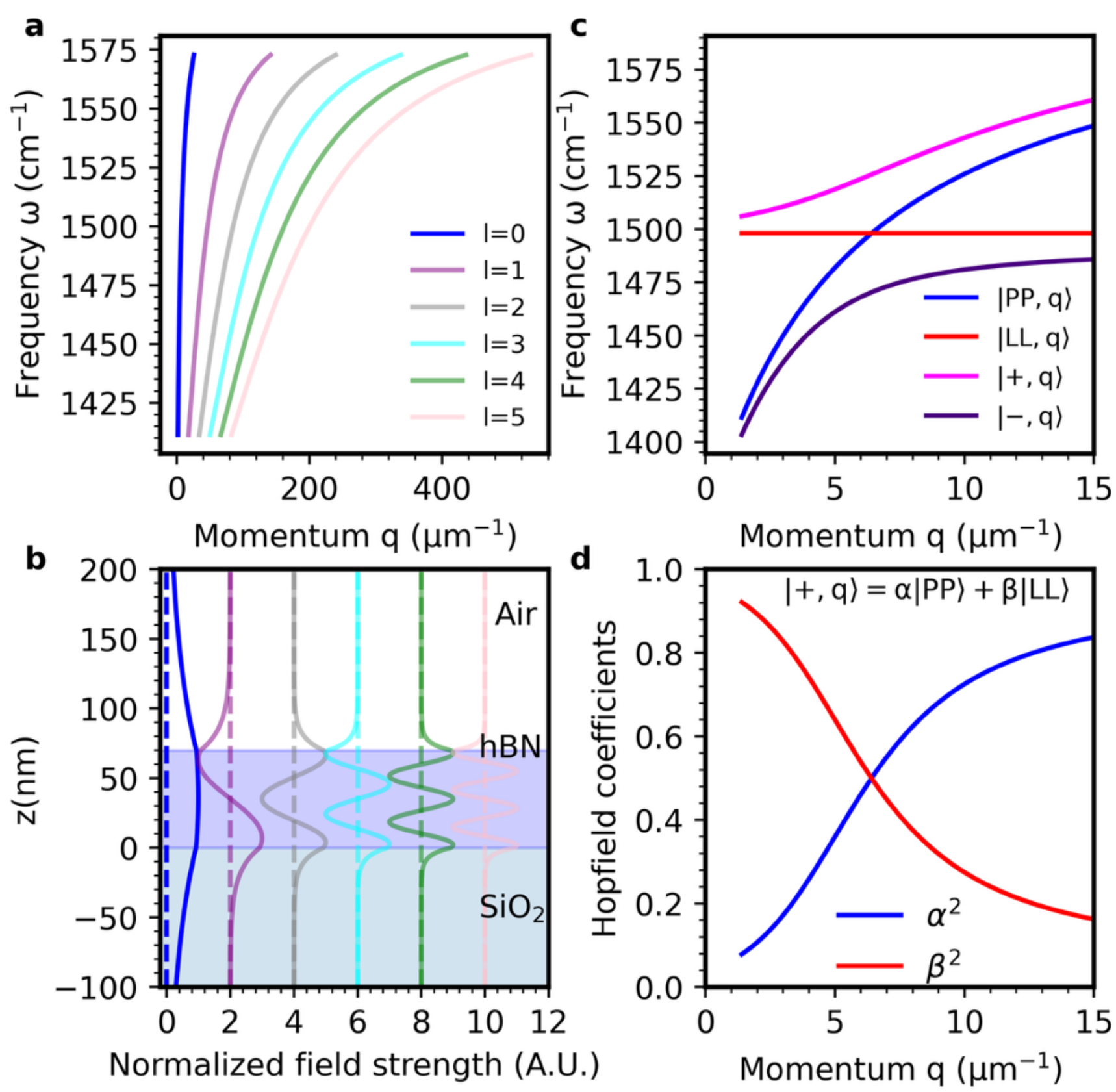

Figure 2. **Quantum electrodynamic modeling for mode coupling between a Landau level resonance and hBN phonon cavity modes.** (a) Theoretically calculated cavity mode dispersions in the type-II branches in a 70-nm-thick hBN slab (without graphene). The index l denotes the order of the cavity mode. (b) Normalized in-plane electric field of hBN cavity modes in the vertical direction as a function of height at $1500\ cm^{-1}$. (c) QED-calculated dispersion relationship in a ($d = 71\ nm, N = 1$) MC. The blue curve corresponds to the zeroth-order cavity mode in hBN. The red curve corresponds to the Landau level resonance. The magenta and the purple curves correspond to the upper and lower branches of the hybridized mode. (d) Momentum-dependent Hopfield coefficients of the upper branch. The blue curve ($\alpha^2$) and red curve ($\beta^2$) represent the contributions from the hBN cavity mode and the ILT, respectively.

## Towards ultrastrong coupling in multilayer cavities

To establish a quantitative baseline for mode coupling in MCs, we first examine the hybridization between the fundamental hBN cavity mode (l = 0) and the inter-Landau-level transitions (ILTs) at $B = 3.2$ T. Here, we compare complex reflectance calculations with the analytical prediction of Eq. (3) to guide future experiments. These simulations employ a weak incident light field to probe the intrinsic linear response of the system without perturbing its properties.

Figures 3a–c show the calculated imaginary part of the p-polarized reflectance coefficient, Im($r_p$), as a function of frequency and in-plane momentum for MCs of fixed thickness (70 nm) with $N = 1$, 3, and 10. Both the numerical calculations (false-color maps) and the formula in Eq. (3) (dashed curves) demonstrate a systematic enhancement of the mode splitting with increasing $N$, reaching $\Delta > 150\ cm^{-1}$ for $N = 10$. Here, the splitting magnitude significantly exceeds values reported in previously studied hBN-based van der Waals heterostructures [30,60,61], highlighting the enhanced hybridization enabled by the multilayer geometry. For clarity, the overlays in Figs. 3a–c emphasize the avoided crossing of the fundamental $l$ = 0 mode. Higher-order modes occur at larger in-plane momenta and contain internal field nodes (Fig. 2b), making their coupling to the graphene Landau-level resonance highly sensitive to layer position. Supplemental Fig. S2 visualizes this dependence, which is discussed in detail in the following section.

Fig. 3(d) summarizes the splitting magnitude of the l = 0 mode as a function of $N$ for cavity thicknesses ranging from 10 nm (red) to 200 nm (blue) at $B = 3.2$ T. For a fixed number of graphene layers, thinner MCs exhibit larger $\Delta$. In the extreme case of large $N/d$, $\Delta$ exceeds the hBN Reststrahlen bandwidth ($\omega_{LO} - \omega_{TO}$). While the simulated spectra largely agree with the prediction of Eq. (3), deviations appear when the splitting becomes large ($\Delta > 150\ cm^{-1}$). As shown in Fig. 3e, all simulation results collapse onto a single curve when plotted as a function of $N/d$, exhibiting an approximate $\sqrt{N/d}$ scaling consistent with Eq. (3). For $N/d > 0.7\ nm^{-1}$, the system enters the ultrastrong-coupling regime, where $\Delta/2$ exceeds 10% of the eigen-frequency $\omega_{ILT}$ [2,3]. However, at large $N/d$, the calculated splitting systematically falls below that of Eq. (3), indicating that the simplified Hamiltonian in Eq. (1) overestimates the coupling strength in densely packed multilayer structures.

This deviation originates from the single-resonance approximation used in Eq. (1), where graphene is modeled as a single Lorentzian oscillator centered at 1500 $cm^{-1}$. Within this approximation, the predicted splitting increases without bound as $N/d \to \infty$. Such divergence represents a formal limit of the zero-thickness-sheet model rather than a physically accessible regime. In reality, graphene hosts multiple Landau-level transitions. As discussed in the supplemental material, the -2→→1 transition can be treated as an isolated oscillator only within a limited spectral window. Outside this range, additional transitions from other LLs contribute to the electromagnetic response and effectively reduce the achievable splitting compared with the single-resonance prediction. Accessing higher-energy transitions, such as the 0→→1 ILT at a stronger magnetic field, would extend the spectral window and increase the attainable splitting. Despite these practical limitations, the multilayer cavity reaches a regime in which splitting exceeds the hBN Reststrahlen bandwidth and enters the ultrastrong coupling regime. These results establish van der Waals multilayer cavities as a scalable and tunable platform for engineering broadband ultrastrong mode coupling.

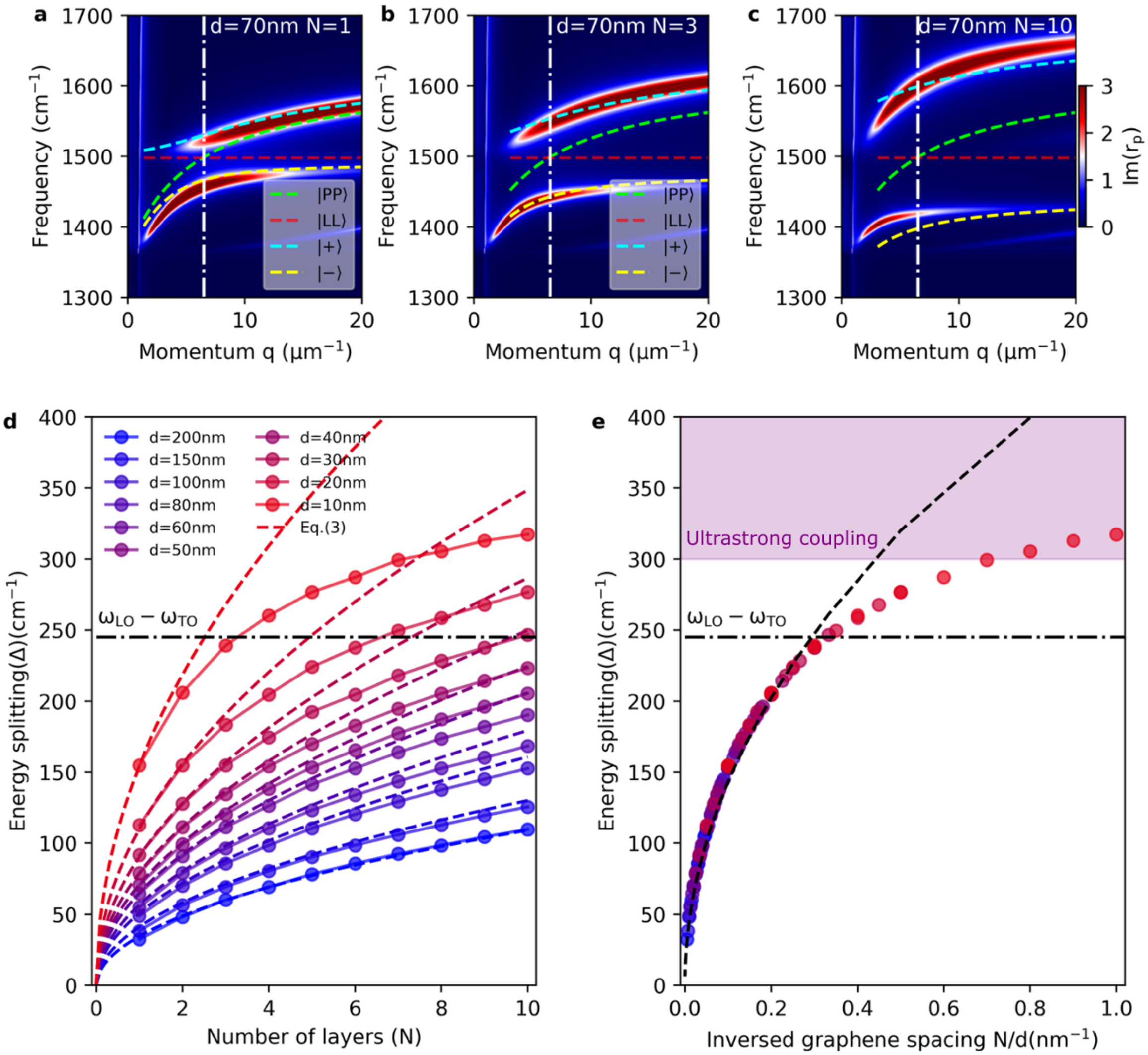

Figure 3. Layer-density-dependent mode hybridization and coupling in graphene/hBN MCs. (a–c) Momentum- and frequency-resolved imaginary part of p-polarized reflectance [Im($r_p$)] for multilayer cavities (MCs) with fixed total thickness (d = 70 nm) and increasing graphene layer count: **(a)** *N* = 1, **(b)** *N* = 3, **(c)** *N* = 10. Green curve: Zeroth-order hBN cavity mode dispersion. Red curve: ILT dispersion. Cyan/yellow curves: Upper/lower branches. White dashed-dotted line: zero-detuning momentum. (d) N-dependent mode splitting energy of MC with varying thickness (from 10 nm to 200 nm). The dashed lines represent the prediction of Eq. (3). The black dash-dotted line represents the width of the Reststrahlen band of hBN. (e) Mode-splitting energy of MC as a function of inverse graphene spacing. The mode-splitting energy, extracted from MCs with different thicknesses, exhibits universal trends in terms of the graphene spacing. The black dashed line is Eq. (3). The shaded purple regime highlights the ultrastrong coupling regime where $\frac{\Delta}{2} > 0.1\omega_{ILT}$ .

## Geometry-dependent coupling of higher-order modes.

For higher-order phonon cavity modes (l > 0), the coupling strength and mode-splitting magnitude $\Delta_l$ become sensitive to the vertical placement of graphene layers within the cavity unit cells [57–59,62–65]. As indicated by Eq. (2), the coupling strength vanishes when a graphene layer coincides with a node of the electric field. This spatial dependence enables selective coupling between the LL resonance and higher-order cavity modes through precise control of layer positioning. Fig. 4 illustrates this principle for an MC ($d = 70$ nm, $N = 2$), where each graphene layer can be placed at an arbitrary height within the hBN slab.

Figures 4a–c present the calculated mode splitting magnitude $\Delta_\mathrm{l}$ as a function of the vertical coordinates of the graphene layers within the full hBN slab ($z_1, z_2$) for l = 0, 1, and 2 modes. The two layer positions are illustrated in Fig. 4d. Line cuts at fixed $z_2$ are shown in the top panels. The fundamental mode (l = 0) exhibits an almost position-independent gap $\Delta_0 \approx 63\ cm^{-1}$, reflecting the nearly uniform electric field profile. In contrast, the l = 1 mode shows vanishing $\Delta_1$ when the graphene layers are located at the midpoints of the slab (red cross in Fig. 4b), corresponding to the field node. Similarly, the l = 2 mode exhibits $\Delta_2 = 0$ when graphene layers are positioned at the trisection points (green cross in Fig. 4c), where the electric field also vanishes. For all mode orders, the splitting is maximized when the graphene layers are positioned near the top or bottom interfaces of the hBN slab, where the electric field amplitude is largest due to boundary conditions and the evanescent character of the confined cavity modes.

The mode selectivity predicted by Eq. (2) is corroborated by numerical reflectance calculations. Figure 4d shows the three representative graphene-layer arrangements marked by the blue, red, and green crosses in Figs. 4a–c. The horizontal black lines indicate the graphene positions, while the colored curves show the vertical electric-field profiles of the $l$ = 0, 1, and 2 cavity modes across the hBN slab. Figures 4e–g show the corresponding calculated Im($r_p$) maps. When graphene layers lie away from the field nodes, clear avoided crossings appear for all cavity modes (Fig. 4e). In contrast, the splitting of the l = 1 and l = 2 modes vanishes when graphene layers coincide with their respective field nodes (Figs. 4f and 4g). This pronounced spatial dependence, absent from effective medium theory (Supplemental Material), shows that coupling depends on both spectral resonance and the spatial structure of the electromagnetic mode profiles. Together, these results validate the QED framework and establish layer-position engineering as a powerful strategy for controlling light-matter interactions in van der Waals heterostructures.

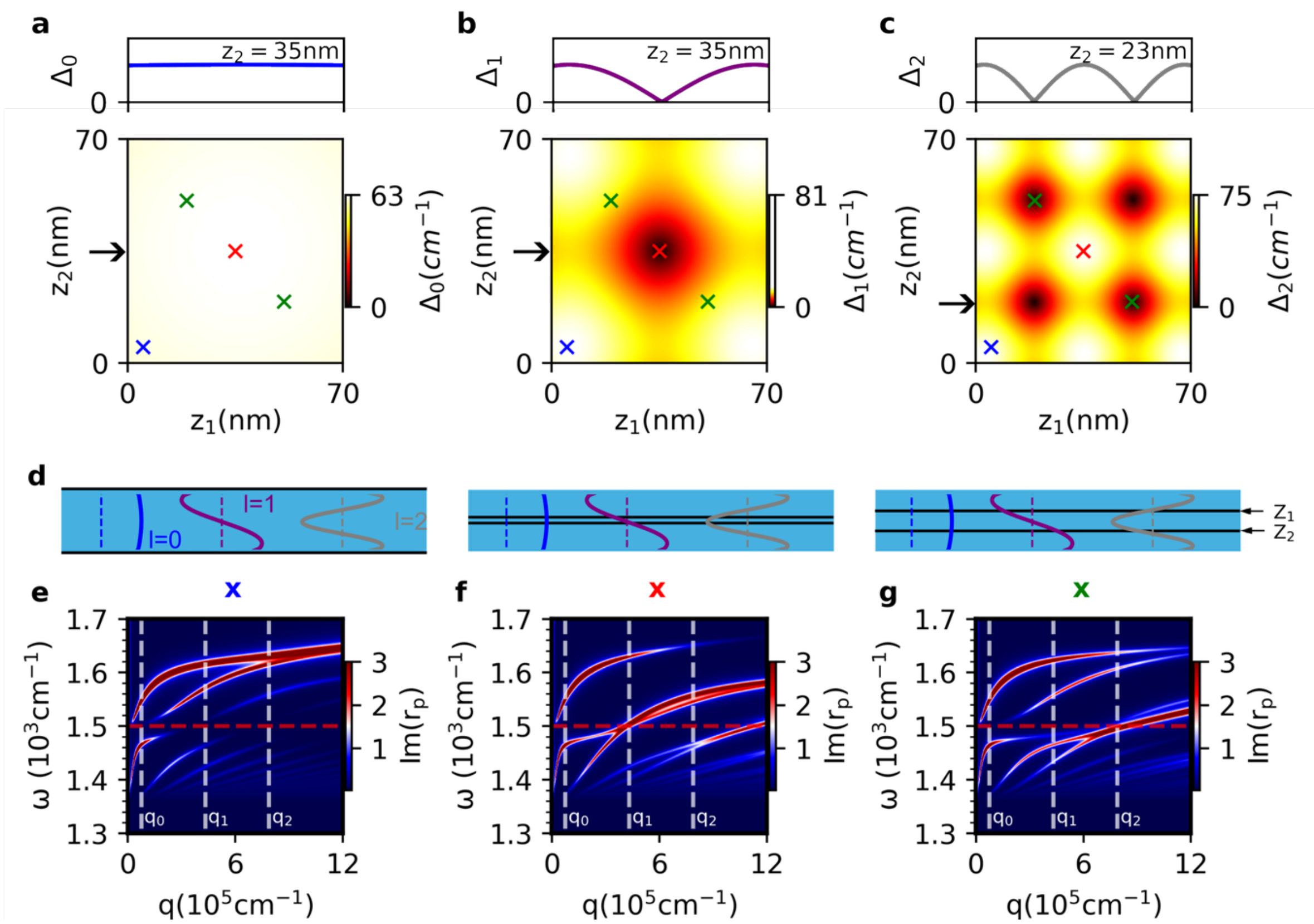

Figure 4. Depth-engineered selective coupling of cavity modes in a two-unit-cell graphene/hBN heterostructure. (a–c) Bottom panels: Calculated mode-splitting magnitude for cavity modes with indices $l = 0, 1, 2$ respectively, in an MC (d = 70 nm, N = 2). The mode-splitting magnitudes are functions of $z_1$ and $z_2$, which denote positions of the two graphene layers relative to hBN. Blue, red, and green crosses mark three representative layer arrangements illustrated in (d). Top panels: Mode-splitting magnitude versus the position of one graphene layer ($z_1$) with $z_2$ fixed; the selected $z_2$ is marked by black arrows in the corresponding bottom panels. (d) Illustration of the arrangements of two graphene layers. The horizontal black lines denote the graphene layers, while the colored solid lines represent the electric field profiles of cavity modes, as shown in Fig. 2b. (e–g) Corresponding Im($r_p$) maps for the three MC layer arrangements, plotted as functions of momentum and frequency. While the l = 0 mode exhibits nearly constant mode splitting $\Delta_{l=0}$ regardless of graphene position, the gaps for the l = 1 and l = 2 modes ( $\Delta_{l=1}$ and $\Delta_{l=2}$) vanish when graphene is placed at the midpoint (f) and trisection point (g) within the hBN slab, respectively. The avoided crossing momenta for l = 0, 1, and 2 modes are labeled with gray dashed lines.

## Discussion

The proposed multi-repeat graphene/hBN cavities provide a versatile platform for magnetically tunable strong mode hybridization. A classical transfer-matrix model captures the linear spectra, and the QED Hamiltonian identifies the collective mode basis, layer-density scaling, and limits of the rotating-wave approximation. With matched material response, coupling terms and dissipation, the harmonic quantum and classical descriptions give the same linear spectra. While the present model captures the essential magneto-optical characteristics of the system, several important directions merit further theoretical and experimental investigation.

In the ultrastrong coupling regime, defined here by $\frac{\Delta}{2} \gtrsim 0.1\omega_{LL}$ , the rotating-wave approximation (RWA) becomes inadequate. A treatment beyond the RWA must consistently include counter-rotating terms and any self-polarization or diamagnetic contribution required by the chosen light–matter representation [2,3]. A key open question is whether a quantum phase transition can emerge in QED systems involving cyclotron transitions of Dirac electrons. Such an instability depends sensitively on the balance between diamagnetic and paramagnetic light-matter coupling [66,67]. Proposals based on spatially nonuniform transverse cavity fields require a sufficiently strong finite-momentum paramagnetic response to overcome the diamagnetic contribution [68,69]. The present MC supports large in-plane momenta, although whether an analogous instability criterion can be satisfied remains unresolved. Recent experiments have demonstrated cavity-induced modifications of integer and fractional quantum Hall states, including changes in exchange splitting and activation gaps [10,70]. These observations motivate exploring whether strongly confined fields in vdW hyperbolic cavities can similarly reshape electronic correlations in graphene.

A second important direction concerns the modification of the hBN phonon response at high graphene density. Increasing the graphene layer density enhances dielectric screening, which can renormalize hBN phonons. Previous studies have shown that long-range Coulomb interactions and environmental screening strongly influence the LO-TO splitting and phonon dispersion in hBN [71]. In densely packed multilayer cavities, thinner hBN spacers place graphene layers closer to the phonon field. Strong screening associated with pronounced ILTs may therefore substantially renormalize the effective hBN phonon response.

Another important consideration is the role of nonlocal graphene conductivity. The local conductivity approximation remains valid when the relevant system length scales, such as cavity thickness and inverse coupling momentum, are much larger than the magnetic length $l_B = \sqrt{\hbar/e|B|}$ [72]. At length scales comparable to $l_B$, however, nonlocal corrections become significant and may activate otherwise forbidden ILTs observed in optical measurements [42,56,73]. These effects introduce additional coupling channels between hBN cavity modes and nominally forbidden ILTs, particularly for higher-order modes at large in-plane momenta. Many-body effects, including Fermi-velocity renormalization and magnetoexciton binding energies, are also neglected in the present model [42,74]. While these effects shift the energies of ILTs or bring new resonances, they are not expected to alter the central conclusions regarding collective mode hybridization and scaling behavior.

Thickness disorder is an important consideration for realistic device implementations, as achieving perfectly uniform interlayer spacing between graphene sheets remains experimentally challenging. As shown in Fig. 4, the coupling between ILTs and higher-order hBN cavity modes is highly sensitive to the precise spatial arrangement of the graphene layers. The quantum-material QED framework, together with Eq. (3), accurately captures the coupling to the fundamental hBN cavity mode, which remains relatively robust against thickness disorder. By contrast, higher-order modes are more susceptible to interlayer variations and therefore require precise knowledge of the graphene layer positions for quantitative modeling.

Beyond these practical considerations, Landau level polaritons provide a promising platform for exploring cavity control of quantum geometry. Landau levels form a paradigmatic topological flat-band system, in which the quantum metric and Berry curvature are closely linked to Chern topology [75,76]. Recent predictions of cavity-modified topological moiré minibands and quantum-geometric bands in chiral-cavity graphene raise the broader question: can cavity hybridization also manipulate the quantum geometry of the graphene Landau levels [77,78]?

Finally, the multilayer architectures and quantum-material QED framework developed here are not limited to graphene/hBN multilayer cavities but can be generalized to hybridization between a broad range of resonances in other van der Waals materials. For example, plasmon–magnon coupling in graphene/antiferromagnet systems [79–82] could be substantially enhanced through multilayer engineering. More generally, this approach provides a systematic route toward the strong- and ultrastrong coupling regimes and opens new opportunities for cavity-engineered quantum materials.

## Acknowledgement

### Funding

M.K.L. acknowledges support from the NSF Faculty Early Career Development Program under Grant No. DMR – 2045425. M.K.L. acknowledges the Gordon and Betty Moore Foundation https://doi.org/10.37807/gbmf12258 for supporting the development of chiral polaritonic materials. L.W. acknowledges funding by Empire State Development Corporation of New York.

### Author contributions

S.X., M.L., and D.N.B. conceived the study. S.X. performed the numerical calculations. B.V. established the theoretical framework with input from S.X., R.J., and M.M.F. S.X. analyzed the calculated and simulated results with assistance from B.V. and R.J. S.X., B.V., R.J., M.L., and D.N.B. wrote the manuscript with input from all coauthors. All authors reviewed and approved the final manuscript.

### Competing interests

The authors declare that they have no competing interests.

**Supplemental material for**

**Quantum materials QED with van der Waals crystals**

# Magneto-optical conductivity of graphene

We calculated the infrared magneto-optical conductivity in graphene based on the Kubo formula and obtained analytical results using an energy-independent scattering-rate approximation [83]. Similar approaches have been used and agree well with experimental results in both the terahertz and infrared frequency ranges [56,84,85].

The longitudinal complex conductivity of graphene in magnetic fields can be expressed analytically in a simple form:

$$\begin{aligned}\sigma_{xx}(\omega) = &\frac{e^2 v_F^2 |eB| (\omega + 2i\gamma)}{\pi c i} \\ &\times \sum_{n=0}^{\infty} \left\{ \frac{[f(E_n) - f(E_{n+1})] + [f(E_{-(n+1)}) - f(E_{-n})]}{(E_{n+1} - E_n)^2 - (\omega + 2i\gamma)^2} \times \frac{1}{E_{n+1} - E_n} \right\} \\ &+ \frac{e^2 v_F^2 |eB| (\omega + 2i\gamma)}{\pi c i} \\ &\times \sum_{n=0}^{\infty} \left\{ \frac{[f(E_{-n}) - f(E_{n+1})] + [f(E_{-(n+1)}) - f(E_n)]}{(E_{n+1} + E_n)^2 - (\omega + 2i\gamma)^2} \times \frac{1}{E_{n+1} + E_n} \right\}\end{aligned} \qquad \text{(S1)}$$

Here, we assumed that the scattering rate, which determines the broadening of Landau-level transitions, is independent of frequency and the Landau level index. $f(E_n) = \frac{1}{e^{\frac{E_n - \mu}{k_B T}} + 1}$ is the Fermi-Dirac distribution function and $E_n = sgn(n)\sqrt{\frac{2n v_F^2 |eB|}{c}}$ is the Landau level energy. The accuracy of the calculated magneto-optical conductivity is determined by the number of inter-Landau-level transitions, and smaller magnetic fields require more Landau levels to be included in the calculation. In this paper, 80000 Landau levels are included in the calculation, and the formula reproduces the graphene conductivity in the low-field limit in the energy range we considered in this work. In the calculation, we used $v_F = 1.2 \times 10^6 m/s$ as the Fermi velocity of graphene and we used $\gamma = 4\ meV$ as the scattering rate in graphene.

# Dielectric functions of hexagonal boron nitride

We modeled the dielectric function of hBN based on the previous far-field result [86]. The phonon resonances in the ab plane and along the c axis can be modeled using a Lorentz-oscillator model:

$$\epsilon_i(\omega) = \epsilon_{i,\infty} + \epsilon_{i,\infty}\frac{\omega_{LO,i}^2 - \omega_{TO,i}^2}{\omega_{TO,i}^2 - \omega^2 - i\omega\gamma_i} \quad \text{(S2)}$$

The parameters are summarized in Table S1.

| Optical axis | $\omega_{LO}(cm^{-1})$ | $\omega_{TO}(cm^{-1})$ | $\epsilon_\infty$ | $\gamma(cm^{-1})$ |
|---|---|---|---|---|
| a and b | 1614 | 1360 | 4.9 | 7 |
| c | 825 | 760 | 2.95 | 2 |

Table S1. **Dielectric function of hBN**

We use a spatially local hBN permittivity model, so the wavevector dependence of the polariton modes arises from the electromagnetic boundary problem; intrinsic phonon spatial dispersion is neglected. This approximation requires field variations to be slow on the lattice scale. For atomically thin spacers, thickness-dependent phonon parameters and discrete-layer effects may also matter, so the large-layer-density limit is a model extrapolation rather than a quantitatively validated device prediction.

# Quantum electrodynamic model for graphene/hBN multilayer cavity

In this section, we derive the Hamiltonian in Eq. (1) of the main text. We first discuss the Hamiltonians for graphene electrons in a magnetic field and for hBN phonons and then consider the coupling between them.

## Graphene magnetoexcitons

We choose the Landau gauge with vector potential $A = -By\,\hat{x}$, where B is the out-of-plane magnetic field. In this gauge, a single-particle eigenstate is specified by its Landau-level (LL) index n, the x-component of momentum $p_x$, and an index $\alpha$ that ranges over the four electron flavors specified by spin (up or down) and valley (K or K'); we denote the corresponding creation operator by $c_{n,p_x,\alpha}^\dagger$ . The energy of the nth LL is $\varepsilon_n = sgn(n)\omega_c\sqrt{|n|}$, with $\omega_c = \sqrt{2}v_F/l_B$, where $v_F$ is the graphene Fermi velocity and $l_B = \sqrt{\hbar c/eB}$ is the magnetic length. The graphene Hamiltonian is

$$H_g = \sum_{p_x,\alpha}\sum_{n=-\infty}^{\infty}\varepsilon_n c_{n,p_x,\alpha}^\dagger c_{n,p_x,\alpha} + \frac{1}{2A}\sum_{\mathbf{q}} v_{\mathbf{q}} {:}\,\rho_{\mathbf{q}}\rho_{-\mathbf{q}}\,{:}, \quad \text{(S3)}$$

where A is the area of the system, $v_{\mathbf{q}}$ is the interaction potential and the electron density $\rho_q$ may be expressed as [87]

$$\rho_{\mathbf{q}} = \sum_{n,m=-\infty}^{\infty} \rho_{nm,\mathbf{q}} b^{\dagger}_{nm,\mathbf{q}}, \tag{S4}$$

with

$$\rho_{nm,\mathbf{q}} = \sqrt{\mathcal{N}}\left[ f_{n_<-1,n_>-1,q} + \operatorname{sgn}(nm) f_{n_<,n_>,q}\right], n_< = \min(|n|,|m|), n_> = \max(|n|,|m|),$$

$$f_{n1,n2,\mathbf{q}} = \left(\frac{\left((1+\delta_{0,n1})(1+\delta_{0,n2})\right)}{4}\left(\frac{n_1!}{n_2!}\right)\right)^{\frac{1}{2}} \left[\frac{\left(l_B(q_x - i\,q_y)\right)}{\sqrt{2}}\right]^{n2-n1} L_{n1}^{(n2-n1)}\left(\frac{1}{2} l_B^2 q^2\right) \exp\left(-\frac{1}{4} l_B^2 q^2\right), \tag{S5}$$

where $\mathcal{N} = \sum_{p_x,\alpha} = \frac{2A}{\pi\, l_B^2}$ is the LL degeneracy, $L_n^k(x)$ is the associated Laguerre polynomial, and

$$b^{\dagger}_{nm,\mathbf{q}} = \frac{1}{\sqrt{\mathcal{N}}} \sum_{p_x,\alpha} c^{\dagger}_{n,p_x,\alpha}\, e^{-i\,\mathbf{q}\cdot\mathbf{r}}\, c_{m,p_x,\alpha}, \tag{S6}$$

where $\mathbf{r} = (i\partial/\partial p_x, l_B^2 p_x)$. The operators $b^{\dagger}_{nm,\mathbf{q}} = b_{mn,-\mathbf{q}}$ obey the commutation relation

$$\left[b_{nm,\mathbf{q}}, b^{\dagger}_{n'm',\mathbf{q}'}\right] = \delta_{nn'}\delta_{mm'}\delta_{\mathbf{q}\mathbf{q}'} - \frac{1}{\mathcal{N}}\left[\delta_{mm'} e^{-\frac{i}{2}\mathbf{q}\times\mathbf{q}' l_B^2} \sum_{p_x,\alpha} c^{\dagger}_{n',p_x,\alpha} e^{i(\mathbf{q}-\mathbf{q}')\cdot\mathbf{r}} c_{n,p_x,\alpha} + \delta_{nn'} e^{\frac{i}{2}\mathbf{q}\times\mathbf{q}' l_B^2} \sum_{p_x,\alpha} \left(e^{i(\mathbf{q}-\mathbf{q}')\cdot\mathbf{r}} c_{m',p_x,\alpha}\right) c^{\dagger}_{m,p_x,\alpha}\right]. \tag{S7}$$

We consider the case where LLs with index $n \leq 0$ are filled and those with $n > 0$ are empty. For $n, n' > 0$ and $m, m' \leq 0$, the two terms in brackets on the RHS of Eq. (S7) vanish in the noninteracting ground state. Although electron-electron interactions induce electron-hole pair excitations above the noninteracting ground state, we expect that they have only a weak effect on the inter-LL transitions (ILTs) in the long-wavelength limit $q l_B \to 0$. Therefore, we neglect the additional terms in Eq. (S7). In this approximation, the operators $b^{\dagger}_{nm,\mathbf{q}}$ with $n > 0$ and $m \leq 0$ obey the bosonic commutation relations $\left[b_{nm,\mathbf{q}}, b^{\dagger}_{n'm',\mathbf{q}'}\right] = \delta_{nn'}\delta_{mm'}\delta_{\mathbf{q}\mathbf{q}'}$ and $\left[b^{\dagger}_{nm,\mathbf{q}}, b^{\dagger}_{n'm',\mathbf{q}'}\right] = \left[b_{nm,\mathbf{q}}, b_{n'm',\mathbf{q}'}\right] = 0$. Therefore, $b^{\dagger}_{nm,\mathbf{q}}$ is identified as the creation operator for an ILT, or magnetoexciton. The Hamiltonian for the magnetoexcitons is

$$H_{\text{ex}} = \sum_{\mathbf{q}} \sum_{n>0,m\leq 0} \hbar\omega^{\text{ex}}_{nm,\mathbf{q}} b^{\dagger}_{nm,\mathbf{q}} b_{nm,\mathbf{q}}, \tag{S8}$$

with $\hbar\omega^{\text{ex}}_{nm,\mathbf{q}}$ the ILT energy. Although $\hbar\omega^{\text{ex}}_{nm,\mathbf{q}}$ contains many-body corrections due to electron-electron interactions, this is a relatively small effect [6]. For simplicity, in our calculations we neglect such corrections and take $\hbar\omega^{\text{ex}}_{nm,\mathbf{q}} = \varepsilon_n - \varepsilon_m$.

## hBN hyperbolic waveguide mode

We now discuss the hBN hyperbolic polariton modes. Let $z$ be the coordinate along the optical axis of the hBN slab of thickness $d$, with the hBN-substrate interface at $z = 0$. We denote the hBN dielectric functions perpendicular and parallel to the optical axis by $\epsilon_\perp$ and $\epsilon_\parallel$, respectively, the substrate dielectric constant by $\epsilon_s$ and the dielectric constant of the medium above hBN by $\epsilon_a$. Therefore, this system may be described by a $z$-dependent dielectric matrix $\boldsymbol{\epsilon}(z) = \mathrm{diag}[\epsilon_x(z), \epsilon_x(z), \epsilon_z(z)]$, where $\epsilon_x = \epsilon_z = \epsilon_s$ for $z < 0$; $\epsilon_x = \epsilon_\perp, \epsilon_z = \epsilon_\parallel$ for $0 \le z \le d$; $\epsilon_x = \epsilon_z = \epsilon_a$ for $z > d$. The polariton modes are obtained by solving Maxwell's equations with this dielectric matrix, which depends on frequency $\omega$. At a given in-plane momentum $\mathbf{q}$, we assume the quasistatic limit $q \gg \omega/c$, in which case the polaritons consist mostly of phonons with a negligible photon component. The problem reduces to solving

$$[\partial_z \epsilon_z(z) \partial_z - \epsilon_x(z) q^2]\Phi = 0 \tag{S9}$$

for the electric potential $\Phi$, with boundary conditions $\Phi(z \to \pm\infty) = 0$. Solving this eigenvalue problem, we obtain the quantization condition [44]

$$q = -\frac{\psi}{d}(\phi_a + \phi_s + \pi l), l = 0,1,2, \dots \tag{S10}$$

where $\psi = -i\frac{\sqrt{\epsilon_\parallel}}{\sqrt{\epsilon_\perp}}$. The electric potential is

$$\Phi_{l,\mathbf{q}}(z) = \Phi_{0,l,\mathbf{q}} \begin{cases} \cos\phi_s e^{qz}, & z < 0 \\ \cos(k_e^z z + \phi_s), & 0 \le z \le d \\ (-1)^l \cos\phi_a e^{-q(z-d)}, & z > d \end{cases} \tag{S11}$$

where

$$k_e^z = iq\frac{\sqrt{\epsilon_\perp}}{\sqrt{\epsilon_\parallel}} = \frac{q}{\psi} \tag{S12}$$

is the $z$-component of momentum of the mode in hBN, and, for $j = s, a$,

$$\phi_j = \arctan\left(\frac{\epsilon_j}{\epsilon_\perp \psi}\right), \tag{S13}$$

with $2\phi_j$ being the reflection phase shift at the interface between hBN and the other medium. The expression for $q(\omega)$ in Eq. (S10) may be inverted to find the hyperbolic phonon polariton dispersion $\omega_{l,\mathbf{q}}$. We denote creation operators for the phonon modes by $a_{l,\mathbf{q}}^\dagger$, in terms of which the Hamiltonian is

$$H_{\text{ph}} = \sum_{\mathbf{q}} \hbar\omega_{l,\mathbf{q}} \left(a_{l,\mathbf{q}}^{\dagger} a_{l,\mathbf{q}} + \frac{1}{2}\right). \tag{S14}$$

The electric potential produced by excitation of the phonon modes is proportional to the ion displacements and may be expanded as

$$\Phi_{\mathbf{q}}(z) = \sum_{l} \Phi_{l,\mathbf{q}}(z)\left(a_{l,\mathbf{q}} + a_{l,-\mathbf{q}}^{\dagger}\right). \tag{S15}$$

It remains to determine the normalization factors $\Phi_{0,l,\mathbf{q}}$. A simple way to do this is through the expression for electromagnetic energy in a dispersive medium:

$$\mathcal{E} = \frac{A}{8\pi}\left\{\int_0^d dz \left[\frac{d}{d\omega}(\omega\epsilon_{\perp})|\mathbf{E}_{\perp}|^2 + \frac{d}{d\omega}\left(\omega\epsilon_{\parallel}\right)\left|\mathbf{E}_{\parallel}\right|^2\right] + \int_{-\infty}^{0} dz\epsilon_s|\mathbf{E}|^2 + \int_d^{\infty} dz\epsilon_a|\mathbf{E}|^2\right\}. \tag{S16}$$

In this expression, we treat the electric field $\mathbf{E}$ as an operator with components $\mathbf{E}_{\perp} = -\partial_z\Phi_{\mathbf{q}}\hat{\mathbf{z}}$ and $\mathbf{E}_{\parallel} = -i\mathbf{q}\Phi_{\mathbf{q}}$. Consider the state $\left|\Psi_{l,\mathbf{q}}(N_{\text{p}})\right\rangle = \left(N_{\text{p}}!\right)^{-1/2}\left(a_{l,\mathbf{q}}^{\dagger}\right)^{N_{\text{p}}}|0\rangle$, which has $N_{\text{p}}$ phonons in the mode $(l,\mathbf{q})$. We impose the condition

$$\left\langle\Psi_{l,\mathbf{q}}(N_{\text{p}})\middle|\mathcal{E}\middle|\Psi_{l,\mathbf{q}}(N_{\text{p}})\right\rangle = \left\langle\Psi_{l,\mathbf{q}}(N_{\text{p}})\middle|H_{\text{ph}}\middle|\Psi_{l,\mathbf{q}}(N_{\text{p}})\right\rangle = \left(N_{\text{p}} + \frac{1}{2}\right)\hbar\omega_{l,\mathbf{q}}. \tag{S17}$$

From Eqs. (S11), (S16), and (S17), we find

$$\Phi_{0,l,\mathbf{q}} = \frac{1}{q}\left(\frac{8\pi\hbar}{dA}\right)^{1/2}\left(\{1-(2k_e^z d)^{-1}[\sin(2\phi_s)+\sin(2\phi_a)]\}\frac{d\epsilon_{\perp}}{d\omega}\right)^{-1/2}\Bigg|_{\omega=\omega_{l,\mathbf{q}}}. \tag{S18}$$

Here we assume $\omega_{l,\mathbf{q}}$ to be in the upper Reststrahlen band; hence, we neglect the weak frequency dependence of $\epsilon_{\parallel}$.

## Landau phonon polaritons

We consider $N$ graphene layers at vertical positions $z_1, \dots, z_N$ in the hBN slab. The electron density operator in the $i$th layer is denoted $\rho_{\mathbf{q},i}$. The electron-phonon interaction Hamiltonian is

$$H_{\text{int}} = \sum_{\mathbf{q}} \sum_{i=1}^{N} e\Phi_{\mathbf{q}}(z_i)\rho_{-\mathbf{q},i}. \tag{S19}$$

The total Hamiltonian is $H = H_{\text{ph}} + H_{\text{ex}} + H_{\text{int}}$. Explicitly,

$$H=\sum_{\mathbf{q}}\Bigg[\sum_{l}\hbar\omega_{l,\mathbf{q}}a_{l,\mathbf{q}}^{\dagger}a_{l,\mathbf{q}}+\sum_{i=1}^{N}\sum_{n>0,m\leq 0}\hbar\omega_{nm,\mathbf{q}}^{\mathrm{ex}}b_{i,nm,\mathbf{q}}^{\dagger}b_{i,nm,\mathbf{q}}$$
$$+\sum_{i=1}^{N}\sum_{l}\sum_{n>0,m\leq 0}e\Phi_{l,\mathbf{q}}(z_i)\rho_{nm,-\mathbf{q}}\left(a_{l,\mathbf{q}}+a_{l,-\mathbf{q}}^{\dagger}\right)\left(b_{i,nm,-\mathbf{q}}+b_{i,nm,\mathbf{q}}^{\dagger}\right)\Bigg]. \quad \text{(S20)}$$

For simplicity, we have ignored the electrostatic coupling between magnetoexcitons in different graphene layers, which has only a weak effect on their dispersion. Suppose the $(l,\mathbf{q})$ phonon mode becomes resonant with the two degenerate optically allowed ILTs $-(n+1)\to n$ and $-n\to n+1$, e.g., $-1\to 2$ and $-2\to 1$. We denote the two corresponding creation operators for these ILTs as $b_{i,1,\mathbf{q}}^{\dagger}$ and $b_{i,2,\mathbf{q}}^{\dagger}$, and their frequency as $\omega_{\mathrm{ILT}}$. From Eq. (S5), $\rho_{-(n+1),n,\mathbf{q}}\propto q$ in the long-wavelength limit $ql_B\to 0$, so these ILTs have a finite in-plane transition dipole moment $\mu_{\mathrm{ILT}}=\lim_{\mathbf{q}\to 0} ie\rho_{-(n+1),n,\mathbf{q}}/q$, which couples to the longitudinal electric field produced by the phonons $E_{l,q}(z)=-\mathrm{i}q\Phi_{l,\mathbf{q}}(z)$. Keeping only resonant terms and using the leading ILT dipole matrix element in the limit $ql_B\to 0$, the Hamiltonian reduces to

$$H=\sum_{\mathbf{q}}\Bigg[\hbar\omega_{l,\mathbf{q}}a_{l,\mathbf{q}}^{\dagger}a_{l,\mathbf{q}}+\sum_{i=1}^{N}\sum_{j=1}^{2}\hbar\omega_{\mathrm{ILT}}b_{i,j,\mathbf{q}}^{\dagger}b_{i,j,\mathbf{q}}$$
$$+\sum_{i=1}^{N}\sum_{j=1}^{2}e\left(E_{l,q}(z_i)\mu_{\mathrm{ILT}}a_{l,\mathbf{q}}b_{i,j,\mathbf{q}}^{\dagger}+\text{ H.c. }\right)\Bigg]. \quad \text{(S21)}$$

It may be diagonalized by a suitable linear transformation of the operators. At the zero-detuning wavevector $q_l$, defined by $\omega_{l,q_l}=\omega_{ILT}$, the upper and lower polariton branches have a frequency gap

$$\Delta_l=\frac{1}{\hbar}\left[8\sum_{i=1}^{N}\left|eE_{l,q_l}(z_i)\mu_{\mathrm{ILT}}\right|^2\right]^{1/2}. \quad \text{(S22)}$$

From Eqs. (S5), (S11), and (S18), we have the formula

$$\Delta_l=\mu\left[\frac{\omega}{2\hbar d_{\mathrm{eff}}}\sum_{i=1}^{N}\mathcal{E}_{l,q_l}{}^2(z_i)\right]^{1/2},\qquad \mu=\frac{e}{\sqrt{2\pi}}\left(\sqrt{n+1}-\sqrt{n}\right),$$
$$d_{\mathrm{eff}}=\frac{\omega}{32\pi}\frac{d\epsilon_{\perp}}{d\omega}d,$$
$$\mathcal{E}_{l,q_l}(z)=\sqrt{2}\{1-(2k_e^z d)^{-1}[\sin(2\phi_a)+\sin(2\phi_s)]\}^{-1/2}\cos(k_e^z z+\phi_s). \quad \text{(S23)}$$

where the dielectric functions and effective confinement length are evaluated at the resonant wavevector and frequency $\omega = \omega_{l,q_l} = \omega^{\mathrm{ex}}_{nm,\mathbf{q}}$. Here $q = q_l,\ k_e^z = q_l/\psi,\ \omega = \omega_{ILT}$; $\mathcal{E}_{l,q_l}(z)$ is the dimensionless mode profile normalized by $\frac{1}{d}\int_0^d \mathcal{E}_{l,q_l}{}^2(z)\, dz = 1$. For large $N$, we have $\sum_{i=1}^{N} \mathcal{E}_{l,q_l}{}^2(z_i) \approx N$ and the formula becomes

$$\Delta_l = \mu\sqrt{\frac{\omega N}{2\hbar d_{\mathrm{eff}}}}. \tag{S24}$$

## Effective medium theory description

In this section, we employ the effective medium theory to approximate the electromagnetic response of the proposed structure. For any nanomaterial with a small but finite thickness, the propagating surface mode is determined by its dielectric function and the surrounding dielectric environment. In the case of a thin metal ($\epsilon$) encapsulated in dielectrics ($\epsilon_1$), the momentum $q$ of the even mode is given by [44]: $\coth\frac{d\sqrt{q^2-\epsilon\omega^2/c^2}}{2} = -\frac{\epsilon\sqrt{q^2-\omega^2/c^2}}{\epsilon_1\sqrt{q^2-\epsilon\omega^2/c^2}}$

The propagating mode exists at the poles of the response function. In other words, the electron system becomes "unstable" under infinitesimal external perturbation and the charge density responds to the external field in a collective out-of-phase fashion. For a type-II hyperbolic mode at 6.17–7.35 $\mu m$ (1620 ~ 1360 cm$^{-1}$) in hBN, its momentum is determined by the asymptotic angle $\psi = -i\frac{\sqrt{\epsilon_\parallel}}{\sqrt{\epsilon_\perp}}$. Here, $\epsilon_\parallel$ ($\epsilon_\perp$) refers to the dielectric function along (perpendicular to) the c-axis and perpendicular (parallel) to the sample surface:

$$\tilde{q} = -\frac{\psi}{d}\left[\arctan\frac{\epsilon_{air}}{\epsilon_\perp\psi} + \arctan\frac{\epsilon_{sub}}{\epsilon_\perp\psi} + \pi l\right].$$

Analogous to the scenario with an ordinary metal, the existence of the type-II hyperbolic mode relies on the negative real part of the in-plane component of the dielectric function, which is induced by the hBN phonon. When the magneto-exciton polariton of graphene aligns with the frequency range of the hBN phonon, the strong oscillator strength of the graphene mode renormalizes the overall in-plane response of the composite stack from metallic to dielectric—a transition from a negative in-plane dielectric function to a positive one turns on the polariton gap.

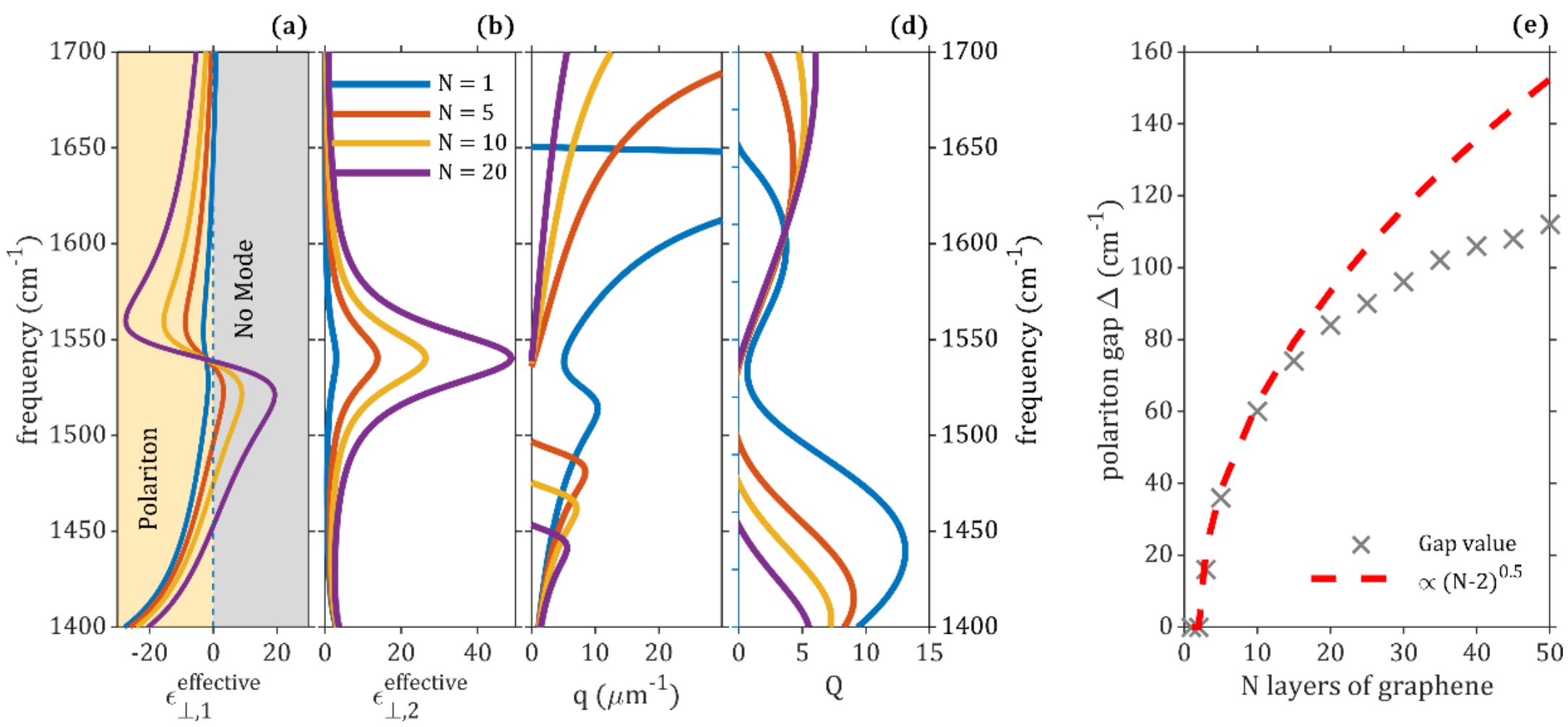

Figure S1. Effective medium calculation of Gr/hBN metastructure. The real and imaginary parts of the in-plane (perpendicular to the c-axis of boron nitride) effective dielectric function are shown in panels (a) and (b). In panels (c) and (d), we illustrate the momentum and the quality factor using the effective dielectric function and the formula. In panel (e), we exhibit the layer-number dependence of the real polariton gap.

To quantitatively capture this behavior, we utilized an effective in-plane dielectric function to approximate the behavior of the polariton (Fig. S1(a) and S1(b)):

$$\epsilon_{\perp}^{\text{effective}} = \frac{\epsilon_{Gr} d_{Gr} + \epsilon_{\perp} d_{hBN}}{d_{Gr} + d_{hBN}}.$$

Here, $d_{Gr}$ is assigned to be 0.6 nm. We further assume that the effective out-of-plane dielectric response differs negligibly from that of bare boron nitride. When N = 1, $\epsilon_{\perp,1}^{\text{effective}}$ develops a kink at the resonance frequency of the graphene magneto-exciton (Fig. S1(a)). Below (above) the kink, the value is closer to (farther away from) 0, predicting a relatively higher (lower) mode momentum illustrated in Fig. S1(c). The second effect of the magnetoexciton resonance is the additional loss of the hybrid mode, indicated by the increased imaginary part of the effective dielectric function (Fig. S1(b)). The additional loss is shown more directly in the quality factor (Fig. S1(d)). When additional graphene layers (N = 5–20) are incorporated into the structure, the real part of the effective dielectric function becomes positive. As a result, the polariton mode is completely suppressed, and a full polariton gap emerges, as evidenced by the absence of momentum (Fig. S1(c)) and quality factor (Fig. S1(d)). Our calculations indicate that beyond N = 2, the mode can be fully suppressed. The gap size up to N = 50 is summarized in Fig. S1(e). The simulated value gradually saturates above 100 cm$^{-1}$. The initial increase of the gap size can be compared to the $(N-2)^{0.5}$ behavior.

## Location-dependent coupling to higher-order cavity modes

Higher-order hBN cavity modes occur at larger in-plane momenta and exhibit multiple electric-field nodes across the hBN thickness. Their coupling to a graphene Landau-level resonance therefore depends not only on spectral detuning but also on the graphene position z, which determines the spatial overlap with the cavity field.

We evaluated this dependence for a single graphene sheet placed at z = 0, 5, …, 70 nm in a 70-nm hBN slab on 285-nm SiO2/Si. Graphene conductivity was estimated at B = 3.225 T, T = 150 K, and μ = 0 eV, and the transfer-matrix calculations covered 1300–1700 cm−1 and in-plane momenta up to $2.0 \times 10^6\ cm^{-1}$.

The higher-q branch morphology varies strongly with z, reflecting the position-dependent overlap between graphene and the cavity-mode fields. This location dependence means that the visibility and splitting of higher-order anticrossings cannot be interpreted without specifying the graphene position.

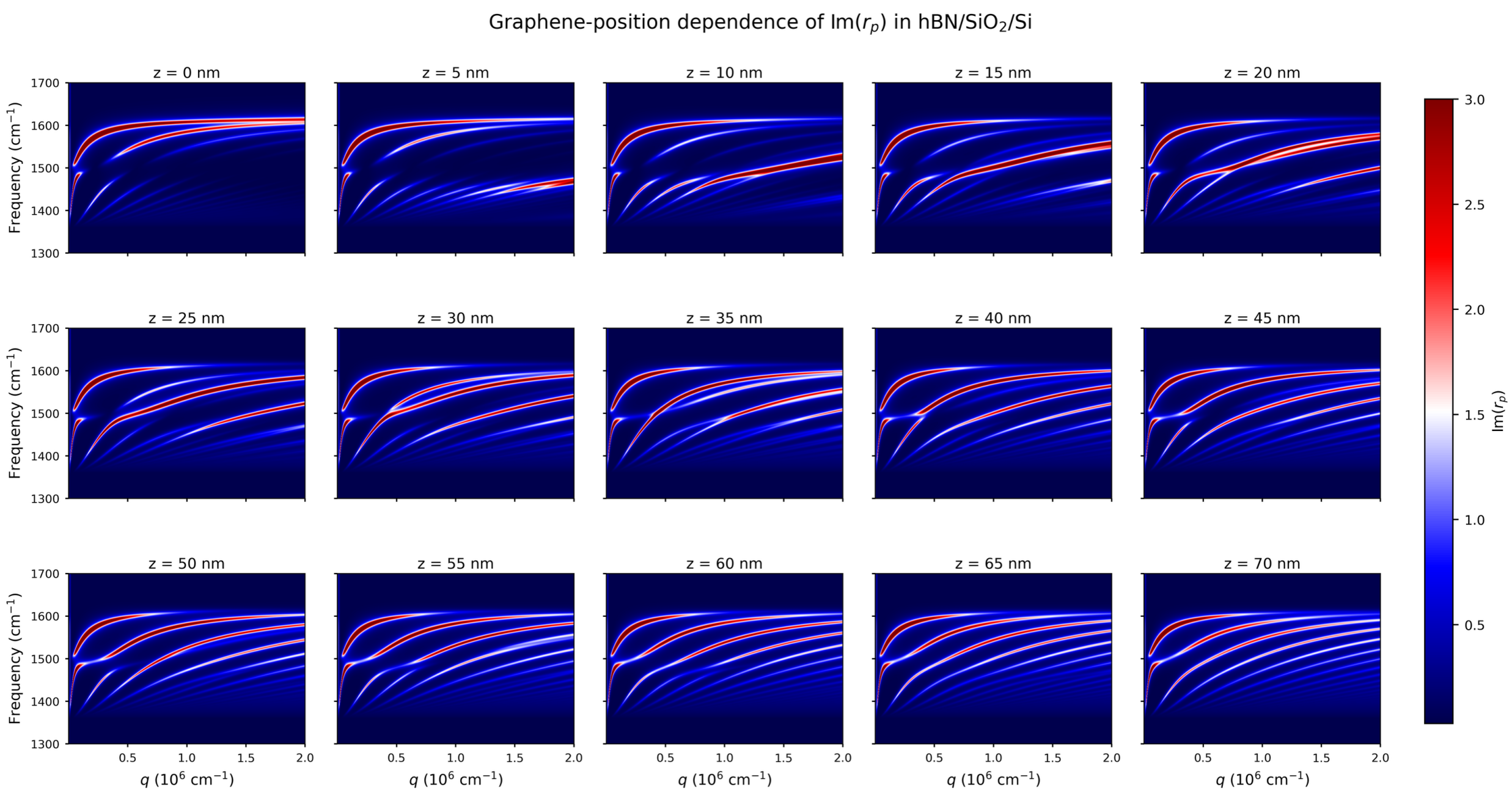

Figure S2. Position-dependent hybridization of higher-order hBN cavity modes. Calculated $Im(r_p)$ for a single graphene sheet at z = 0, 5, …, 70 nm within a 70-nm hBN slab on 285-nm SiO2/Si. The coordinate z is measured from the top hBN interface; z = 0 and 70 nm therefore place graphene at the top and bottom interfaces, respectively. The panels show $Im(r_p)$ as a function of in-plane momentum and frequency at B = 3.225 T, T = 150 K, and μ = 0 eV. Higher-q polariton branches display position-dependent avoided-crossing-like bending and splitting near the Landau-level resonance.

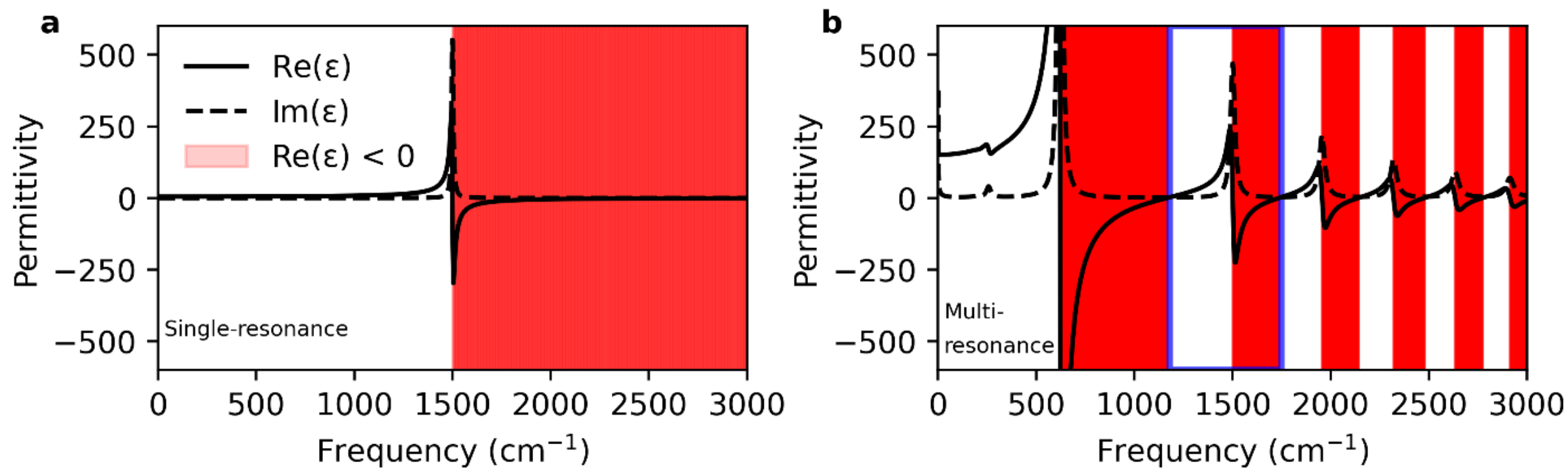

Figure S3. Single-resonance model versus multi-resonance model. (a) Permittivity of a Lorentz oscillator with a single resonance, as considered in the QED model. The red shaded area marks the spectral range where Re($\epsilon$) is negative. (b) Two-dimensional permittivity of charge-neutral graphene under a magnetic field of B = 3.2 T. The blue box highlights the spectral range where the single-resonance QED model remains valid.